\documentclass[twocolumn,showpacs,epsf,psfig]{revtex4-1}
\usepackage{amssymb}
\usepackage{epsfig}
\DeclareMathAlphabet{\pazocal}{OMS}{zplm}{m}{n} 
\usepackage{graphicx}
\usepackage{color}
\usepackage{bm}
\usepackage[normalem]{ulem}   
\usepackage{soul}
\usepackage{amsmath}

\begin{document}
\title{Electric field tuning of the anomalous Hall effect at oxide interfaces} 

\author{Sayantika Bhowal} 
\email{bhowals@missouri.edu}
\affiliation{Department of Physics \& Astronomy, University of Missouri, Columbia, MO 65211, USA}
\author{S. Satpathy}
\affiliation{Department of Physics \& Astronomy, University of Missouri, Columbia, MO 65211, USA}

\begin{abstract}

We show that the anomalous Hall effect (AHE)    
at a magnetic interface with strong spin-orbit coupling 
can be tuned  
 with an external electric field. 
 By altering the strength of the inversion symmetry breaking, 
 the electric field changes the Rashba interaction,
  which in turn modifies the magnitude of the Berry curvature, the central quantity in determining the anomalous Hall conductivity (AHC). 
  The effect is illustrated with a square lattice model, which yields a quadratic dependence of  the AHC for small electric fields.
Explicit density-functional calculations were performed for the 
recently grown iridate  interface, viz.,  the (SrIrO$_3$)$_1$/(SrMnO$_3$)$_1$ (001) structure,
both with and without an electric field. 
The effect may be potentially useful in spintronics applications.



\end{abstract}

\date{\today}	


\maketitle 


The anomalous Hall effect (AHE) occurs in solids with broken time-reversal symmetry, such as the ferromagnets, as a result of the spin-orbit
coupling (SOC). 
Although the effect was noticed in the original work of Hall himself \cite{Hall1, Hall2}, 
the explanation of the phenomenon came from the seminal paper of Karplus and Luttinger \cite{KL},
where they identified
 the anomalous contribution 
to arise from the SOC, which results in  the left-right asymmetry in the scattering of the spin-polarized electrons. 
 Currently, there is a considerable interest on the AHE from a technological point of view because of potential applications in 
 spintronics such as for magnetic sensors and memory devices \cite{Gerber}.    
 
 The interface between 3$d$ anti-ferromagnetic insulator SrMnO$_3$ (SMO) \cite{Takeda} and 5$d$ paramagnetic metal SrIrO$_3$ SIO \cite{Zhao, Zeb, Cao} is one of the notable examples among several attempts \cite{Nichols, Matsuno, Wei, Pang, Ryan} to engineer the electronic and magnetic properties at the 3$d$-5$d$ interfaces, where the strong coupling 
is achieved by the charge transfer from SIO to the SMO side \cite{Nichols,Okamoto}, as sketched in Fig. \ref{fig1}. 
 This results in  electron doped SMO and hole doped SIO, both of which become ferromagnetic.
 The ferromagnetism at the interface in turn gives rise to the AHE, which has been measured for the short-period superlattices of SIO/SMO \cite{Nichols}.
 
 In this paper, we show that the AHE can be tuned by an external electric field
 by modifying the strength of the Rashba interaction. 
 The effect is illustrated using general arguments as well as from density-functional calculations of 
 the AHC for a specific interface structure (SIO)$_1$/(SMO)$_1$, which has already been experimentally grown.
 Such a perovskite hetero-structure is a good candidate for the electric field control of the Rashba effect  
  \cite{Shanavas}, providing an excellent platform for the manipulation of the AHE.    
%

%

\begin{figure}[htb]
\centering
\includegraphics[scale=0.30]{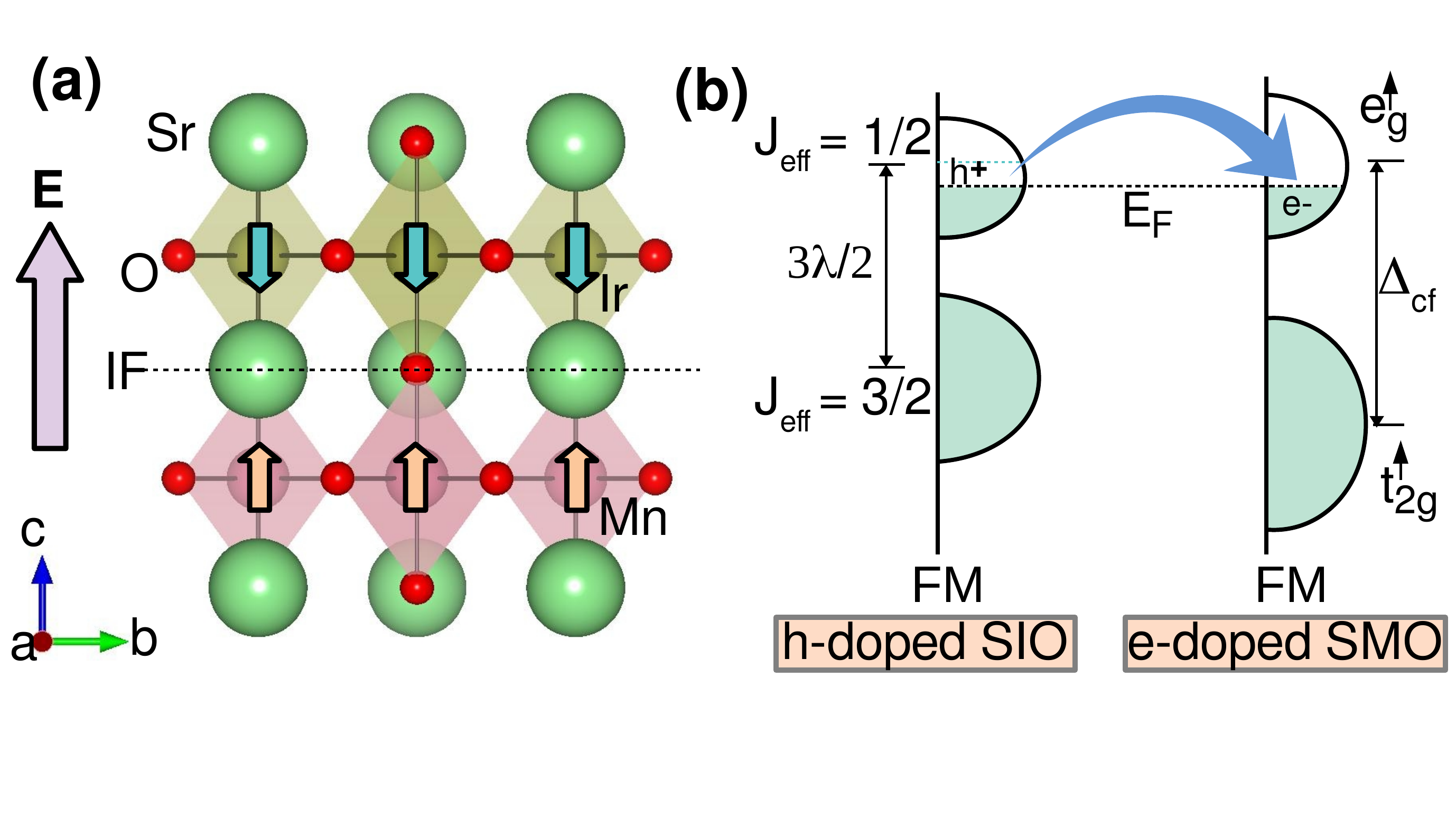}
\caption {
Electronic and magnetic  structure of the (SIO)$_1$/(SMO)$_1$ interface, 
both sides consisting of a single layer each,
 considered here as a specific example for the tuning of the AHC. The  charge transfer across the interface leads to electron or hole doping, which in turn results in a ferromagnetic system on either side, leading to  an anomalous Hall effect.
%
} 
\label{fig1} 
\end{figure}

\begin{figure}[h]     
\centering
\includegraphics[scale=0.21]{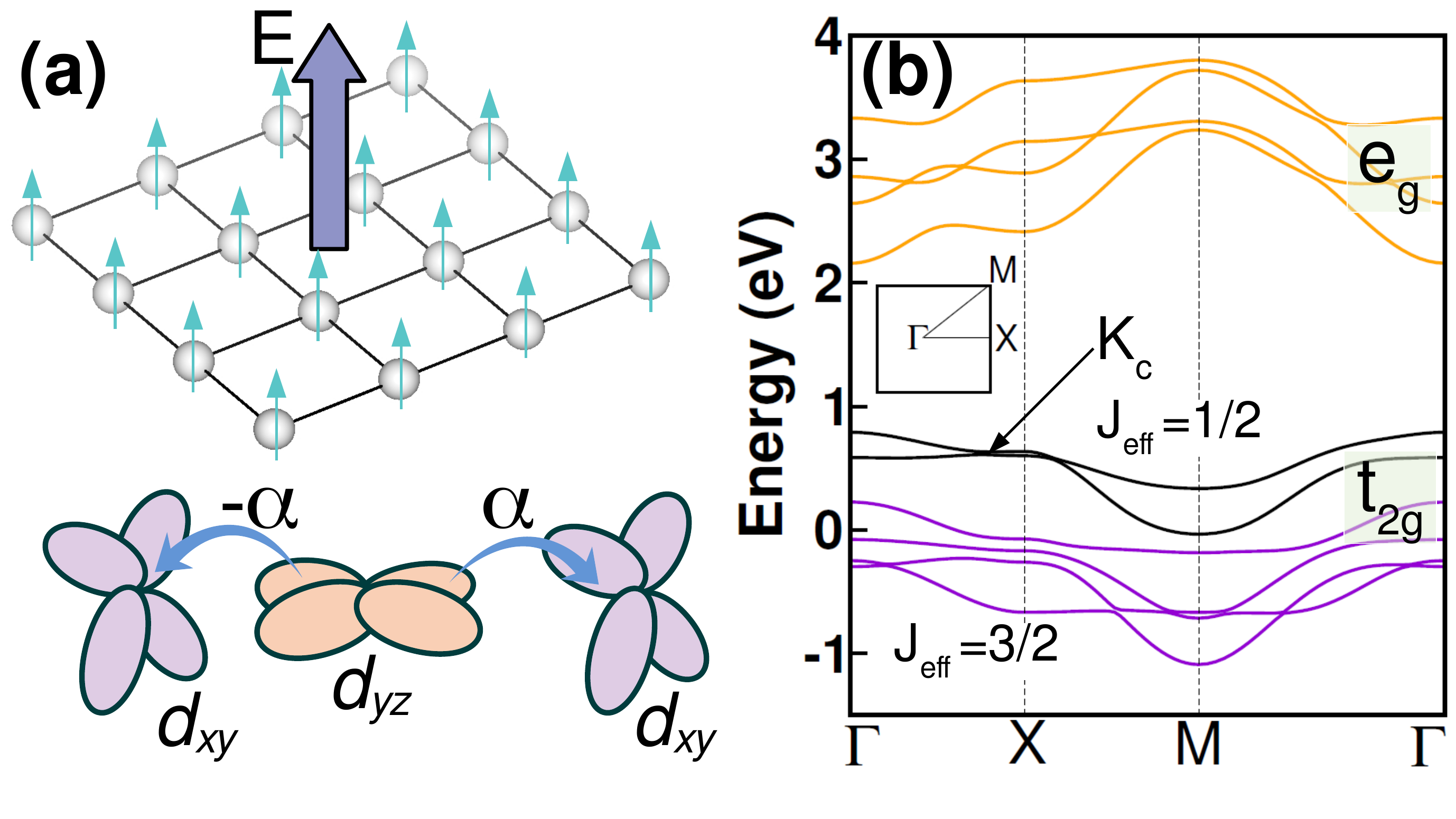}     
\includegraphics[scale=0.25]{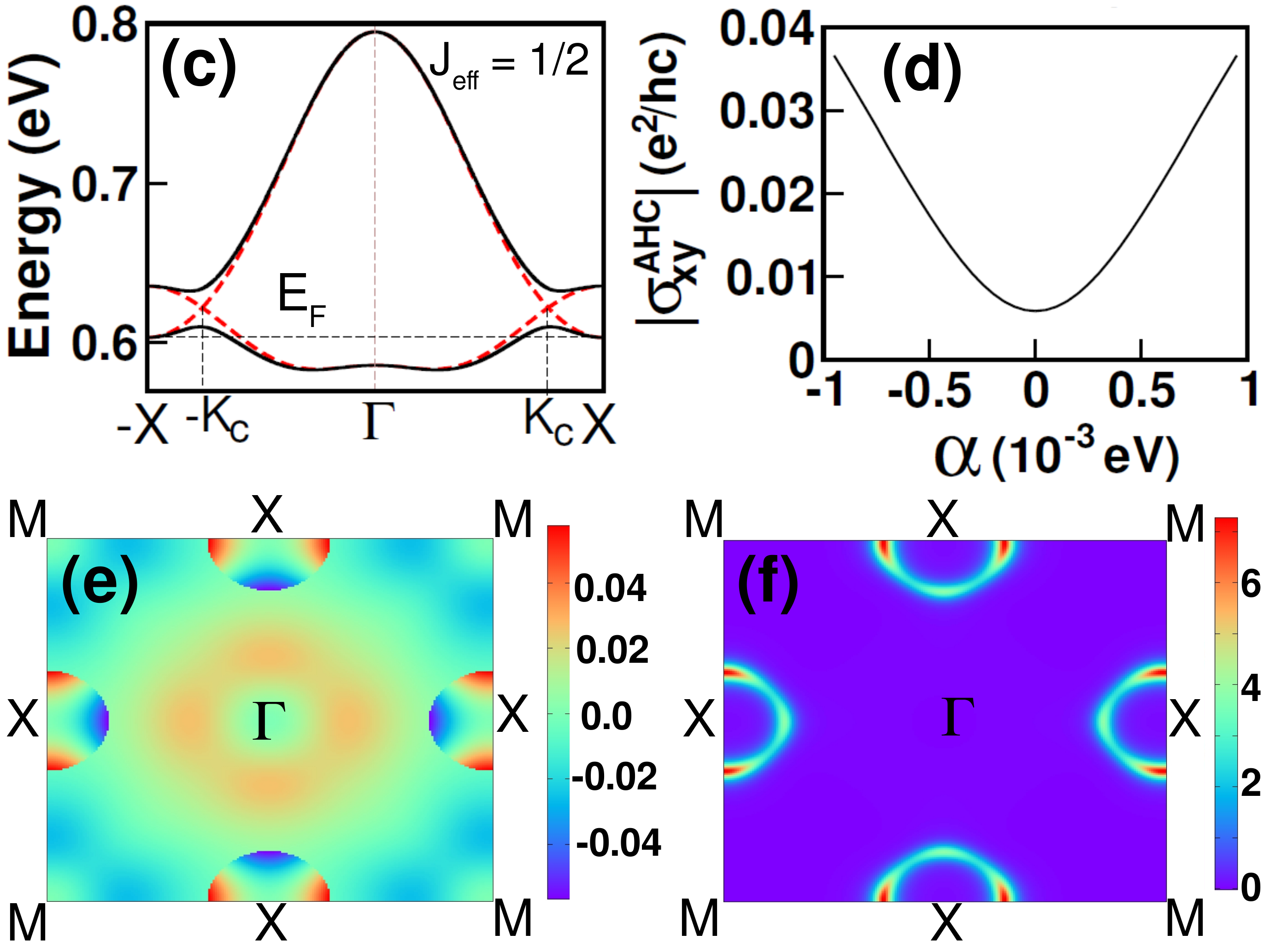}
\caption{ Illustration of the electric field dependence of the Berry curvature  and AHC,
computed from Eqs. \ref{AHC} and \ref{kubo}, for the square-lattice TB model. 
(a) The square lattice  and the electric-field induced TB hopping integral $\alpha$. 
(b) The TB band structure with both   large  crystal field $\Delta_{\rm cf}$  and   SOC parameter $\lambda$,
which is relevant for SIO, where the $J_{\rm eff} = 1/2$ state is partially occupied.
(c) Dispersion of the $J_{\rm eff} = 1/2$ bands with and without an electric field 
(black and red lines, respectively).
(d) Computed AHC for small electric fields, 
characterized by the parameter $\alpha$,
indicating the $\sigma_{xy}^{AHC} \propto |E|^2$ dependence. The Fermi energy $E_F$ corresponded to the 
electron concentration $n_e = 0.9$  in the $J_{\rm eff} = 1/2$ bands.
(e) and (f)  
Berry curvature $\Omega^z_n ({\vec  k})$ (in units of \AA$^2$) for the lower $J_{\rm eff} = 1/2$ band without and with the electric field, 
respectively. 
$\Omega^z_n ({\vec  k})$ is large
near a crossing point $K_c$ (here close to X) and has a dominant contribution to AHC. 
The TB parameters are: $V_\sigma= -0.2 $ eV (1NN),  -0.1 eV (2NN), $V_\sigma/V_\pi =-1.85 $, $J_{ex} = 0.5$ eV,
$\alpha = \beta = \gamma = 0.01$ eV (0 if $E = 0$), and $\Delta_{cf} = 3 $ eV.
} 
\label{fig2}
\end{figure}

To illustrate the effect of the electric field on AHE, 
consider the motion of electrons in a simplified tight-binding (TB) model of a ferromagnetic square lattice [Fig. \ref{fig2} (a)], 
relevant for the transition metal atoms
on either side of the interface. 
The  Hamiltonian is
\begin{eqnarray}       \label{TBM} 
{\cal H}= & & {\cal H}_{kin} +{\cal H}_{ex} + {\cal H}_{SOC} +{\cal H}_E                  \nonumber \\
   = & & \sum_{i\mu\sigma, j\nu\sigma} t^{\mu \nu}_{ij} c_{i\mu \sigma}^\dagger c_{j\nu \sigma} 
- J_{ex} \sum_{i\mu} \sum_{\sigma,\sigma^\prime} c_{i\mu \sigma}^\dagger \sigma_{\sigma\sigma^\prime}^z c_{i\mu\sigma^\prime} \nonumber \\
& + & \frac{\lambda}{2} \sum_{\eta} \sum_{\mu\sigma,\nu\sigma^\prime} c_{i\mu \sigma}^\dagger 
L^\eta_{\mu\nu}   \sigma^\eta_{\sigma\sigma^\prime} c_{i\nu \sigma^\prime}     +     {\cal H}_E,   
\end{eqnarray}
 where we consider $d$ electrons,  $c_{i\mu \sigma}^\dagger$  creates an electron at the $i$-th site with spin $\sigma$ and orbital index $\mu$, $t^{\mu \nu}_{ij}$ is the spin dependent hopping between near neighbors, $J_{ex}$ describes the spin splitting of up and down electrons in the ferromagnet, and
 $\lambda \ \vec  L \cdot \vec  S$ is the SOC term.
%
 In the TB model, the electric field induces asymmetry of the orbital lobes, which opens up new inter-orbital hopping channels \cite{Shanavas, Petersen}, that were zero before.
This is incorporated in the final term ${\cal H}_E$, having the same form as ${\cal H}_{kin}$, 
but with new matrix elements $t^{\mu \nu}_{ij}$, viz., 
  $\alpha = \langle xy | {\cal H}_E | yz\rangle_{\hat x} = \langle xy | {\cal H}_E | xz\rangle_{\hat y}$, $\beta = \langle xz | {\cal H}_E | z^2\rangle_{\hat x} =\langle yz | {\cal H}_E | z^2\rangle_{\hat y}$,
 $\gamma = \langle xz | {\cal H}_E | x^2 - y^2\rangle_{\hat x} = \langle x^2 - y^2 | {\cal H}_E  | yz\rangle_{\hat y} $,  which are roughly  proportional to the electric field with the subscript $\hat x$ or $\hat y$ indicating the location of the nearest neighbor.
 %


The electric field breaks the inversion symmetry and leads to a Rashba interaction in the presence of the SOC.
The TB form ${\cal H}_E $
leads \cite{Shanavas} to the equivalent Rashba Hamiltonian in the momentum space \cite{Rashba}
\begin{equation}
{\cal H}_R = \alpha_R ( \vec k \times \vec \sigma )  \cdot \hat z,
\label{Rashba}
\end{equation}
which results in the   linear-k splitting of the 
band structure $\varepsilon_k = \frac{\hbar^2 k^2}{2m} \pm \alpha_R k$, when $J_{ex} = 0$. 
The Rashba coefficients are different for different bands and can be expressed in terms of the  
matrix elements $\alpha$, $\beta$, and $\gamma$, which are roughly proportional to $E$; 
For instance, $\alpha_R \approx 4 \alpha / 3$ for the $J_{\rm eff} = 1/2$ states \cite{Shanavas}. 
%
In 3D continuum, the SOC term
$
 {\cal H}_{SO}=\frac{\hbar^2}{2m^2c^2} (\vec\nabla V \times \vec {k}) \cdot \vec \sigma,
$
with the potential gradient $\vec \nabla V = -E \hat z$,
immediately leads to  the linear field dependence 
$\alpha_R = - \frac{\hbar^2 E}{2m^2c^2}$.
In the solid, the predominant contribution to $\alpha_R$ comes from the electric field near the nucleus
 \cite{Shanavas}, 
but it still increases linearly with the applied field $E$ as seen from the DFT results presented in Fig. \ref{fig-DFT} (e).





The AHC can be computed \cite{ChangNiu} from the  momentum sum of the Berry curvature 
\begin{equation}
  \sigma^{\rm AHC}_{xy}   =     -\frac{e^2}{\hbar} \frac{1}{N_k V_c} \sum_{n \vec  k} \Omega_n^z ({\vec  k}),
  \label{AHC} 
\end{equation}
where the sum is over the occupied states, and the Berry curvature  $\Omega_n^z ({\vec  k})$ for the $n^{\rm th}$ band
can be evaluated using the Kubo formula \cite{Thouless} 
\begin{equation}         \label{kubo}
 \Omega^z_n ({\vec  k}) = - 2 \hbar^2      \sum_{n^\prime \neq n} \frac {{\rm Im} \langle \psi_{n{\vec  k}} | v_x | \psi_{n^\prime{\vec  k}} \rangle  
                        \langle \psi_{n^\prime{\vec  k}} | v_y | \psi_{n{\vec  k}} \rangle} 
                        {(\varepsilon_{n^\prime \vec k}-\varepsilon_{n  \vec k})^2}. 
\end{equation}
Here 
$v_{\eta} =  \hbar^{-1}   \partial H / \partial k_\eta$, 
$V_c$ is the unit cell volume, and $N_k$ is the number of k points used in the BZ sum.
Near a band crossing point close to $E_F$, which we denote by $K_c$ [see Fig. \ref{fig2} (b) and (c)],  the denominator in (\ref{kubo}) becomes small, leading to a large contribution 
to the AHC. For a crossing point deep below $E_F$, the contributions to the AHC from the two crossing bands cancel due to the opposite signs of the matrix elements.

The  computed values of the Berry curvature using these expressions  for the TB model in absence and presence of electric field are shown in Fig. \ref{fig2} (e) and (f) respectively,
from which it is clear that the band crossing points have the dominant contributions to the Berry curvature. 
The calculated AHC for small electric fields, characterized by the field-induced TB parameter $\alpha$ in ${\cal H}_E$,
is shown in Fig. \ref{fig2} (d), 
which indicates the square-law dependence 
$\sigma^{\rm AHC}_{xy}  = \sigma_0 + c E^2  $. 
	The AHC can also be tuned by a gate voltage, which adds carriers to the system. 
	The results obtained for the TB model are summarized in Fig. \ref{fig3},
	indicating the strong dependence of the AHC on the applied electric field,
	characterized by the parameter $\alpha$, as well as the electron concentration $n_e$, which can be modified with the gate voltage.  

\begin{figure}[h]
\centering
\includegraphics[scale=0.20]{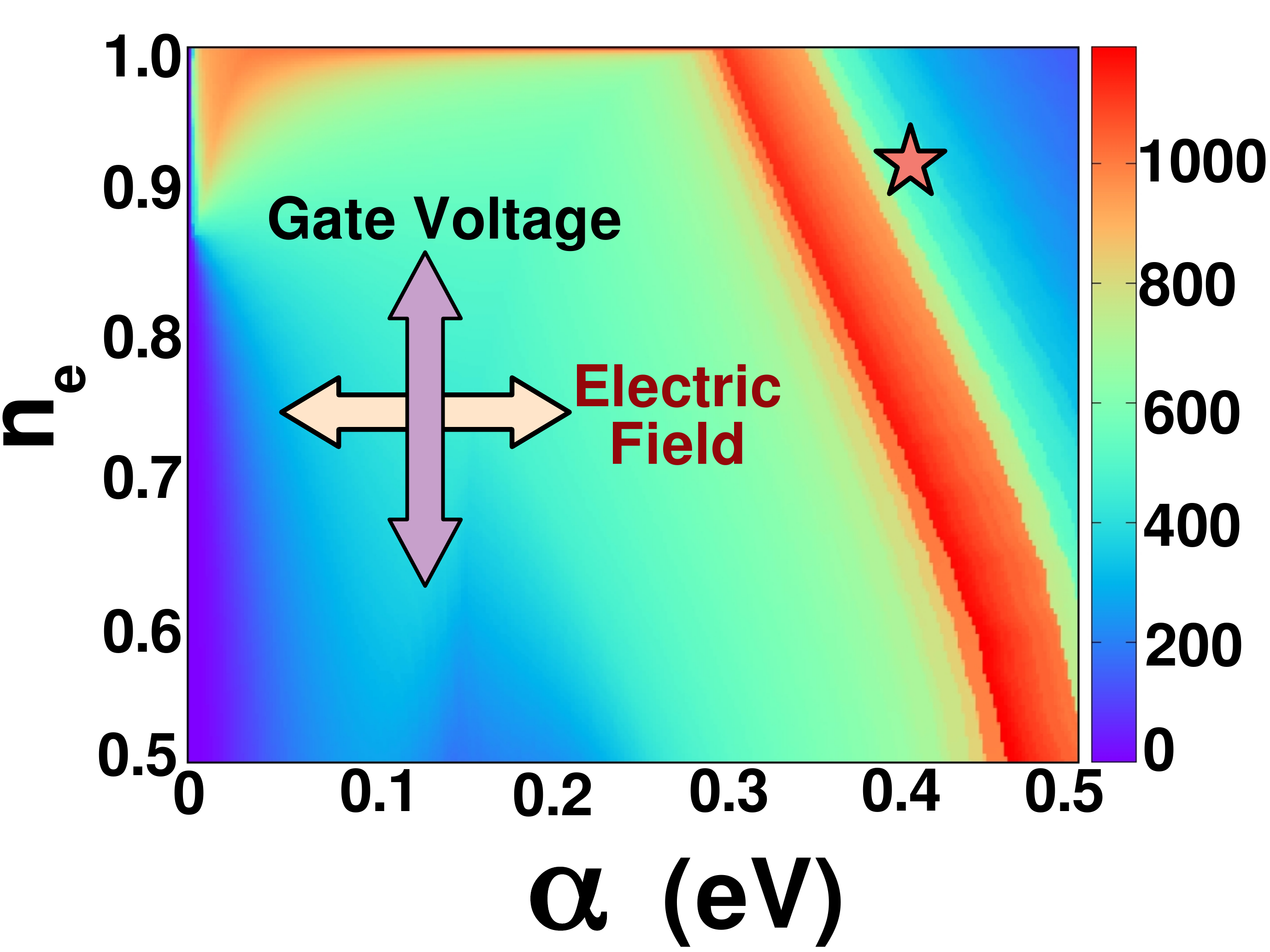}
\caption{ Variation of the AHC with electric field, parametrized by $\alpha$, and the carrier concentration $n_e$
[electrons in the  $J_{\rm eff} = 1/2$ band; see 
Fig. \ref{fig2} (c)], computed for the TB model (Eq. \ref{TBM}). 
The star corresponds to parameters for SIO/SMO and units of AHC are $\Omega^{-1} {\rm cm}^{-1}$. 
} 
\label{fig3}
\end{figure}


The $\sigma_{\rm xy}^{\rm AHC} \propto |E|^2  $ dependence  for small electric fields
can be understood by  
considering the $ 2 \times 2$ Hamiltonian near the crossing point $\vec K_c$
\begin{eqnarray}     \label{H-crossing}  
{\cal H} (\vec q) &= 
\left[
{\begin{array}{*{20}c}
     \eta  q & h_{12}   \\
      h_{12}^* &   -\eta q \\
\end{array} }  \right],
\end{eqnarray} 
where for $E=0$, we have the conical bands $\varepsilon_\pm = \pm \eta q $,
and $h_{12}$ is the electric field dependent term.
Explicitly, we take the crossing point in the $J_{\rm eff} = 1/2$ band, so that the
TB form of ${\cal H}_E$ yields the expression
$h_{12} =   \alpha_R ( \sin k_y + i \sin k_x)$, where $\alpha_R = 4\alpha/3$,
obtained straightforwardly from the Bloch functions corresponding to the  $J_{\rm eff} = 1/2$  wave functions:
$\psi_\pm = (   | yz, \bar \sigma \rangle   \pm i  | xz, \bar \sigma \rangle \pm |xy, \sigma \rangle   ) / \sqrt 3$.
From the eigenvalues of Eq. (\ref{H-crossing}), viz., 
$ \varepsilon_\pm = \pm \sqrt {\eta^2 q^2 + |h_{12}|^2}$,  
and the corresponding wave functions, 
we find the Berry curvature from Eq. (\ref{kubo})
to be
\begin{equation}
\Omega_\pm^z = \pm \frac{\eta}{2}  \times \frac   {-\alpha_1 \alpha_2 q + c_1  \sin \theta - c_2 \cos \theta - \alpha_1 \alpha_2 q \cos (2\theta)}
 {    (\eta^2 q^2  + |h_{12}|^2)^{3/2}    }
\label{model-omega}
\end{equation}
where we have kept the terms to the lowest order in $\vec q \equiv  \vec k - \vec K_c$, so that
$h_{12} = \Delta +\alpha_1 q_y + i \alpha_2 q_x $,
where 
 $\alpha_1 = \alpha_R \cos K_y$, $\alpha_2 = \alpha_R  \cos K_x$,  and $\Delta = \alpha_R (\sin K_y + i \sin K_x)$.
In Eq. \ref{model-omega}, $c_1 = \alpha_2 {\rm Re} (\Delta)$, $c_2 = \alpha_1 {\rm Im} (\Delta)$,  
and the $\pm$ sign refers to the upper and the lower
bands. 
For  $\alpha \ll \eta$, valid for small electric fields,
we immediately find the angle-integrated Berry curvature
to be
\begin{equation}
I_\pm(q) \equiv \int_0^{2\pi} \Omega_\pm^z (q, \theta) \ d \theta = \mp  \frac   {f \alpha_R^2 q \eta }
 {   2 (\eta^2 q^2  + |\Delta|^2)^{3/2}    },
 \label{Iq}
\end{equation}
where $f = \cos K_x \times \cos K_y$. This equation together with Eq. \ref{AHC} clearly shows that $\sigma_{\rm xy}^{\rm AHC} \propto |E|^2  $, since the Rashba coefficient $\alpha_R $ scales as the 
electric field strength. Furthermore, it is clear that $I_\pm(q)$ is sharply peaked close to the band crossing point.
In the square-lattice model, we find the AHC to scale as:
$\sigma^{\rm AHC}_{xy}  = \sigma_0 + c E^2  $ [see Fig. \ref{fig2} (d)], 
where $\sigma_0 \ne 0$ due to the broken time-reversal symmetry. 
Note that this result is valid only for small $E$; For sufficiently large $E$, the bands can realign, 
and    the pre-factor $c$ can get modified as well,  
sometimes even becoming negative,
as seen from the DFT results (Table \ref{conductivity}) for a large positive electric field.  
This is further elaborated in the Supplementary Materials \cite{SM}.

 \begin{table} [h]    
\caption{Total AHC ($\sigma^{\rm AHC}_{\rm xy} = \sigma_{\rm c} + \sigma_{\rm rest}$) and the 
partial contributions, $\sigma_{\rm c}$ from the crossing point $K_c$ in the BZ and the remaining part
$\sigma_{\rm rest}$,
as a function of the applied electric field $E$. 
Note that $\sigma_{\rm rest}$ shows little change with $E$,
while $\sigma_c$ changes significantly, controlling the  electric field behavior of the AHC.}
\centering
 \begin{tabular}{c c c c}
\hline\hline
Electric Field & $\sigma_{\rm rest}$ & $\sigma_{\rm c}$ & $\sigma^{\rm AHC}_{\rm xy}$\\
 (V/\AA)    &    $(\Omega~{\rm cm})^{-1} $ &   $(\Omega~{\rm cm})^{-1} $ & $(\Omega~{\rm cm})^{-1} $\\
\hline
 -0.3 & 23.2 & 13.8 & 37 \\
 -0.05 &  23.3  & 7.7  & 31 \\
  0   &  23.0  & 10.0    & 33 \\
 0.05 &  23.5  & 10.5 & 34 \\ 
 0.3 & 23.0 & -13.7   & 9.3\\  
\hline\hline
\end{tabular} 
\label{conductivity}
\end{table}


 We now turn to the DFT calculations for the (001)   (SIO)$_1$/(SMO)$_1$  slab
 to illustrate 
 the field tuning effect for a real material. 
We used the plane wave methods to solve the DFT equations 
within the GGA+SOC+U approximation \cite{vasp, QE, gga}.
The AHC was calculated by computing the Berry curvatures 
 using the Wannier interpolation approach\ as implemented in the Wannier90 code \cite{w90}.
Further details are given in the Supplementary Materials \cite{SM}.

A key feature of the electronic structure  of the (001) (SIO)$_1$/(SMO)$_1$ interface 
 is the charge transfer \cite{Okamoto, Bhowal} from the spin-orbital entangled $J_{\rm eff} = 1/2$ state on the SIO side to the empty Mn-$e_g$ states on the SMO side [Fig. \ref{fig1} (b)].  The charge transfer is important because it drives both sides ferromagnetic, 
 thereby breaking the time-reversal symmetry, which is an essential ingredient for AHC. 
 The electron-doped SMO becomes ferromagnetic due to the Anderson-Hasegawa-DeGennes double exchange \cite{AH-DEX},
while the hole-doped SIO becomes ferromagnetic due to the Nagaoka physics, 
where in the infinite-U limit, a single doped carrier in the half-filled Hubbard model destroys the anti-ferromagnetic insulating ground state, driving the system into a ferromagnetic metal \cite{Nagaoka}. 

The amount of the charge transfer depends on the exact structure. For the (001)   (SIO)$_1$/(SMO)$_1$  slab,
we find that there is a transfer of about 0.08 $|e|$ across the interface, enough to make both sides ferromagnetic.
  We find the ferromagnetic moments to be 3.12 $\mu_B$ (0.03 $\mu_B$) for spin (orbital) moment for Mn, while for Ir, it is 0.14 $\mu_B$ and 0.08 $\mu_B$, respectively, which are similar to the bulk values. Total energy calculations with constrained spin directions indicate the moments to be aligned along $\hat z$ (normal to the plane) in agreement with the experimental results \cite{Nichols}. 
In order to make contact with the existing experiments, 
we first computed the AHC for the (001) (SIO)$_1$/(SMO)$_1$ superlattice structure. 
The computed value $\sigma_{\rm xy}^{\rm AHC} \approx 26 \  \Omega^{-1}$ cm$^{-1}$  
 is in reasonable agreement with the experimental value of $\sim$ 18 $\Omega^{-1}$ cm$^{-1}$ \cite{Nichols}.


\begin{figure}[tbh]
\centering
\includegraphics[scale=0.32]{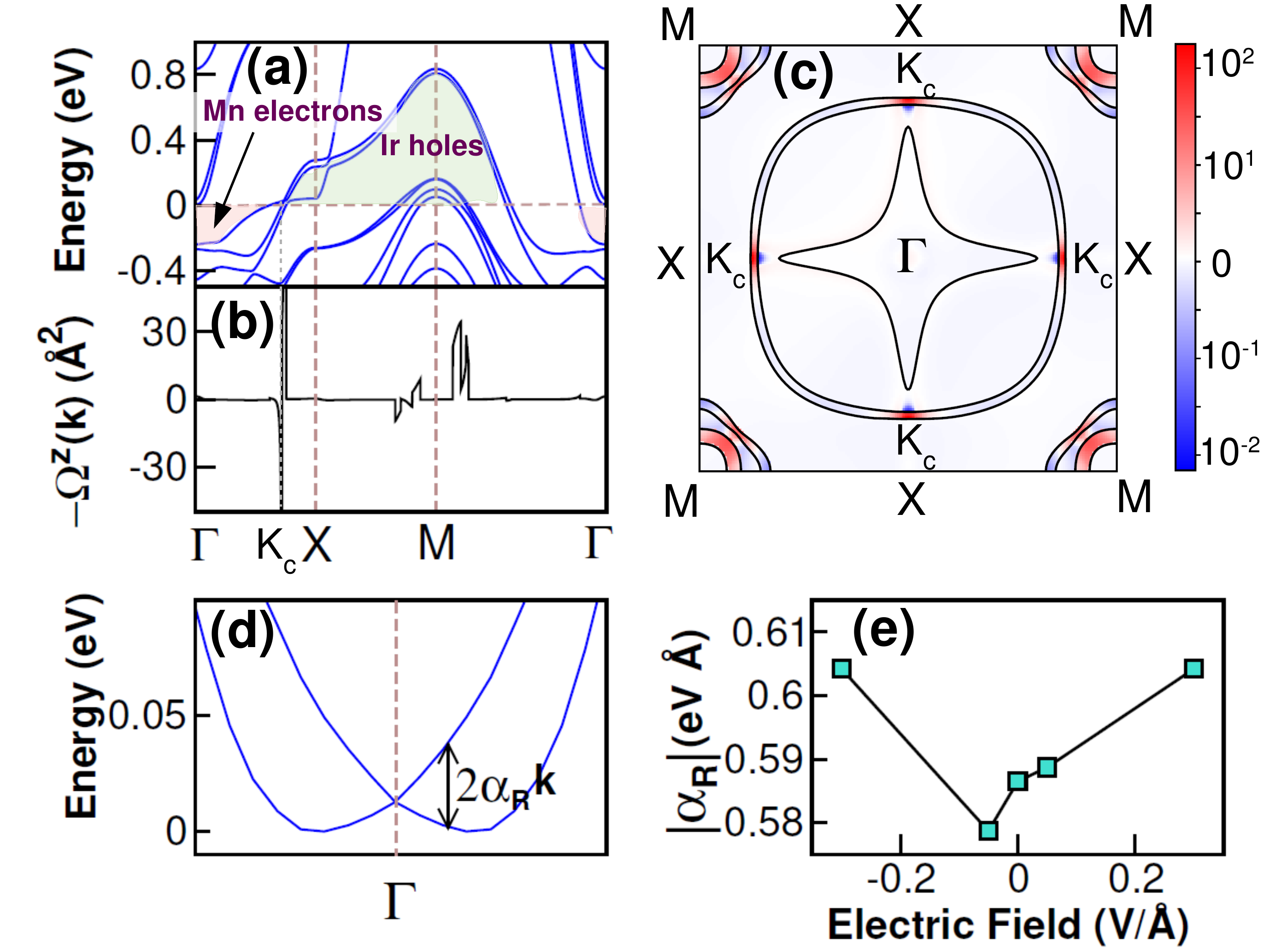}
\caption{ DFT results for the AHC of the (SIO)$_1$/(SMO)$_1$ slab with an applied electric field E. 
(a) Band structure  
for 
E = -0.3 V/\AA, with the colored regions indicating the Ir t$_{2g}$ holes and Mn e$_g$ electrons, and the band crossing point $K_c$.
(b)  The  Berry curvature $ \Omega^z_n ({\vec  k})$ summed over the occupied bands $n$ at each $\vec k$ point
along the specified line in the BZ.
(c) Contours of the same quantity in (b) in the $k_x - k_y$ plane, 
indicating the large contributions from regions around $M$ and
$K_c$ points in the BZ, with the latter providing the dominant contribution to the field dependence of the AHC as discussed in the text. 
(d) Rashba splitting of the DFT band structure for the non-magnetic state, used for extracting $\alpha_R$.
(e)  Variation of the computed $\alpha_R$ with the applied electric field.
}
\label{fig-DFT}
\end{figure}

The typical band structure for the (SIO)$_1$/(SMO)$_1$ is shown in Fig. \ref{fig-DFT} (a), where the Ir holes and the 
Mn electrons are shown, which is consistent with the charge transfer across the interface, as sketched in Fig. \ref{fig1}. It is essential to optimize the crystal structure
for each case in order to take into account the electrostatic screening effect, which reduces the applied field.
There are only subtle changes in the band structure, e.g., around $\vec K_c$, for different electric fields, but the overall band structure remains the same, and there is no substantial change of the charge transfer up to the electric fields we used in the calculations.

As already mentioned, large contributions to the AHC comes from regions in the BZ, 
where both occupied and unoccupied bands occur near the Fermi energy for same $\vec k$, 
which can be seen from the small energy denominator in the Kubo formula (\ref{kubo}). 
As seen from Fig. \ref{fig-DFT} (b) and Table \ref{conductivity},
there are two regions with significant contributions to the AHC,  $\sigma_c$ from the region around the four crossing points $K_c$, 
which strongly varies with the electric field, and the remaining part $\sigma_{\rm rest}$, which remains more or less unaffected
because unlike near $K_c$, the bands change very little at $M$, which is the major contributor to $\sigma_{\rm rest}$.  
The electric field dependence of $\sigma_c \propto \alpha_R^2$ comes from Eq. \ref{Iq}, with $\alpha_R = - \frac{\hbar^2 E}{2m^2c^2}$ in the free particle model as mentioned already. To evaluate this for the solid,
we computed the Rashba coefficient $\alpha_R$ as a function of the electric field  from the linear band splitting $\Delta_k = 2 \alpha_R k$ near the $\Gamma$ point from additional DFT calculations for the non-magnetic structure. 
The results, Fig. \ref{fig-DFT} (d) and (e), show the anticipated linear $E$ dependence of $\alpha_R$. 
Note that for $E = 0$, $\alpha_R$ is significantly large, which can be attributed to an intrinsic electric field $E_0$  that exists
at the interface due to the broken inversion symmetry. 
From the computed $\sigma_{\rm xy}^{\rm AHC}$, we estimate $E_0 \approx 0.6 $ V/\AA. 
Thus, for small electric fields, Eq. \ref{Iq}  yields the result
$\sigma_{\rm xy}^{\rm AHC} \approx  (\sigma_c^0 E_0^{-2}) \times (E_0 + E)^2 + \sigma_{\rm rest}^0$, 
where $\sigma_c^0$ and $\sigma_{\rm rest}^0$ are the contributions for $E = 0$.

So far, we described the electric field tuning via the modification of the Rashba SOC by the applied electric field. 
A second way to alter the AHC is by manipulating the carrier density by a gate voltage. 
This is verified by shifting the Fermi energy in the DFT calculations to a lower value, thereby increasing the Ir-hole concentration. In presence of an electric field E, shifting of Fermi energy downwards by $\Delta \varepsilon_F = - 0.1$ eV enhances the AHC by 15\% to about 38 $\Omega^{-1} {\rm cm}^{-1}$. 
For $\Delta \varepsilon_F = - 0.15$ eV, it is further increased to the value 85 $\Omega^{-1} {\rm cm}^{-1}$.
This offers an additional tool for the electrical manipulation of  the AHC.

In conclusion, we have shown that the anomalous Hall effect at the 3$d$-5$d$ interfaces can be tuned by modifying the Rashba spin-orbit interaction with the application of an external electric field.
The major contribution to the electric-field dependence comes  from the band-crossing points close to the Fermi energy
and varies quadratically for small electric fields. In addition, the AHC can be tuned by manipulating the electron density with a gate voltage. 
We illustrated the results with a ferromagnetic square-lattice model as well as
with density-functional calculations for the recently grown manganite-iridate interface, viz.,
(001) (SIO)$_1$/(SMO)$_1$. 
It would be valuable to develop this effect further, both theoretically and experimentally, 
with an eye towards potential spintronics applications.

We thank the U.S. Department of Energy, Office of Basic Energy Sciences, Division of Materials Sciences and Engineering for financial support under Grant No. DEFG02-00ER45818.


\newpage
\onecolumngrid{
\begin{center}
 \large\bf{Supplementary Materials for \\
Electric field tuning of the anomalous Hall effect at oxide interfaces}
\end{center}
\section{Density-functional methods}
In this section, we will discuss the detail of the electronic structure calculations presented in the paper.  
In order to study the magnetic properties of SrIrO$_3$(SIO)/SrMnO$_3$(SMO), DFT calculations have been performed using the plane-wave based projector augmented wave (PAW) \cite{PAW1,PAW2} method as implemented in the Vienna {\it ab initio} simulation package (VASP) \cite{vasp} within the generalized gradient approximation (GGA) \cite{gga} including Hubbard U \cite{U} and SOC. The magnetic calculations are performed with the unit cell containing two formula units of SMO and SIO where the in-plane lattice parameters ($a = b$) are fixed to the value of experimental lattice constant of the substrate SrTiO$_3$ (3.905 $\times \sqrt{2}$ \AA). The kinetic energy cut-off of the plane wave basis was chosen to be 550 eV. Following the previous report \cite{Okamoto}, all the calculations have been performed using U = 2 eV for Ir and U = 3 eV for Mn-$d$ states respectively.

In order to take into account the electrostatic screening effects, it is important to relax the atomic positions. Therefore, we have relaxed the structure in presence of each of the electric fields using VASP. The atomic relaxations of the slab are carried out in presence and absence of the electric field until the Hellman-Feynman forces on each atom becomes less than 0.01 eV/\AA. For the calculations in presence of electric field, a sawtooth-like  potential (see Fig. \ref{sawtoothPot}) is applied.

The AHC of the superlattice structure and the slabs in presence and absence of electric field are calculated using QUANTUM ESPRESSO and the Wannier interpolation approach \cite{QE, w90}. Self-consistency with magnetization along the (001) direction is achieved using fully relativistic norm-conserving pseudopotentials under the Perdew, Burke, and Ernzerhof generalized-gradient approximation \cite{gga} for all the atoms with a convergence threshold of 10$^{-8}$ Ry. Note that, the magnetic ground state obtained from the previous calculations can be realized in a smaller unit cell with one formula unit of SMO and SIO. This smaller unit cell is used for the calculation of AHC. Using a non-self-consistent (nscf) calculation the {\it ab-initio} wave functions of this ground state are obtained on a regular k-mesh 8$\times$8$\times$4 and 10$\times$10$\times$2 for the superlattice and the slab geometry respectively. The  {\it ab-initio} wave functions thus obtained are used to construct the maximally-localized Wannier functions \cite{MLWF} using the wannier90 code \cite{w90_code}. Some additional empty states are considered in the nscf calculation, that help to localize the Wannier functions. In the disentanglement process, as initial projections, we have chosen 68 and 76 Wannier functions per unit cell for the superlattice and the slab geometry respectively that include the $d$ orbitals of both Mn and Ir and $s$ and $p$ orbitals of O excluding the rest. Accordingly, we have chosen the "inner window" from the bottom of the valence band to an energy slightly above the Fermi level, while the "outer window" includes all the states above that valence band bottom. After the disentanglement is achieved, the wannierisation process is converged to 10$^{-7}$ \AA$^2$ having an average spread less that $\sim$ 1 \AA$^2$ of the Wannier functions. 

\begin{figure}[h]
\centering
\includegraphics[scale=0.40]{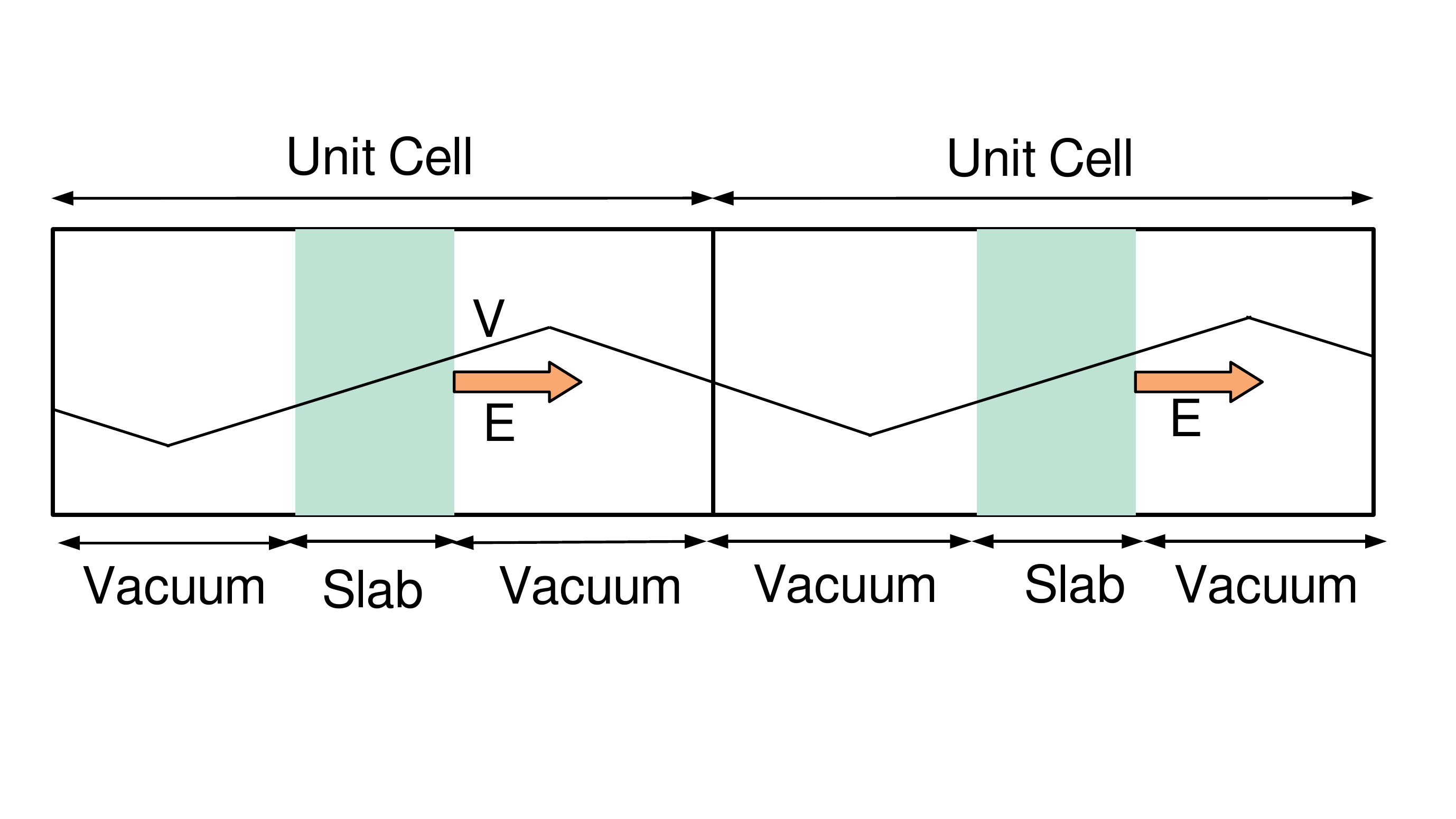}
\caption{The sawtooth-like potential V applied to the (SIO)$_1$(SMO)$_1$ slab.} 
\label{sawtoothPot}
\end{figure}

The AHC is therefore calculated by computing the sum of the Berry curvature of the occupied bands over the Brillouin zone (BZ). The BZ integration of the Berry curvature is done by using a k-mesh of 300$\times$300$\times$150 and 400$\times$400$\times$80 for the superlattice and the slab geometry respectively with an "adaptively refined" mesh \cite{w90} of 7$\times$7$\times$7 when the absolute value of sum of the Berry curvature of the occupied bands at each {\bf k} {\it i.e.}, $\Omega^z ({\bf k})$ is larger than 100 \AA$^2$. The convergence is confirmed by using finer mesh. 

\section{Optimized structure and electrostatic screening} 

As mentioned earlier, the structure of the (SIO)$_1$/(SMO)$_1$ slab is optimized using VASP. The displacements of the different atoms with respect to the ideal structure  is shown in Fig. \ref{fig1} (a). The further displacements of the atoms in presence of an electric field {\bf E} = 0.3 V/\AA~is also shown in \ref{fig1} (b). In the ideal slab, the in-plane lattice constant is fixed to the value of the experimental lattice constant of the substrate SrTiO$_3$ (3.905 \AA)  and the thickness of the SMO and SIO layers are fixed to the corresponding lattice constants of the bulk structures {\it i. e.}, 3.80 \AA~for SMO \cite{Takeda} and 3.94 \AA~ for SIO \cite{Zhao, Longo}. The slabs are separated in the $z$-direction by $\sim$ 12 \AA~of vacuum. 
 
\begin{figure}[h]
\centering
\includegraphics[scale=0.30]{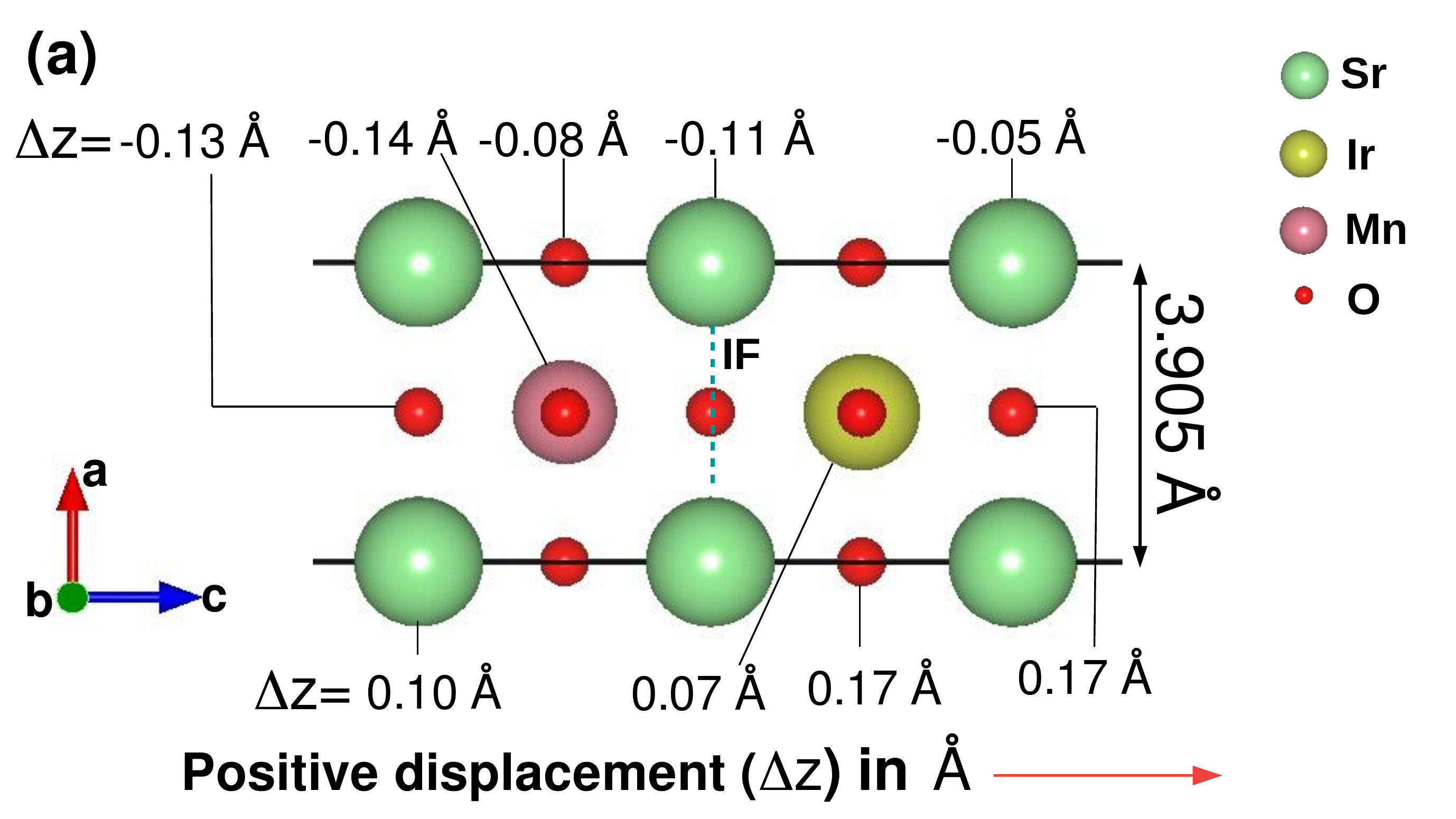}
\includegraphics[scale=0.30]{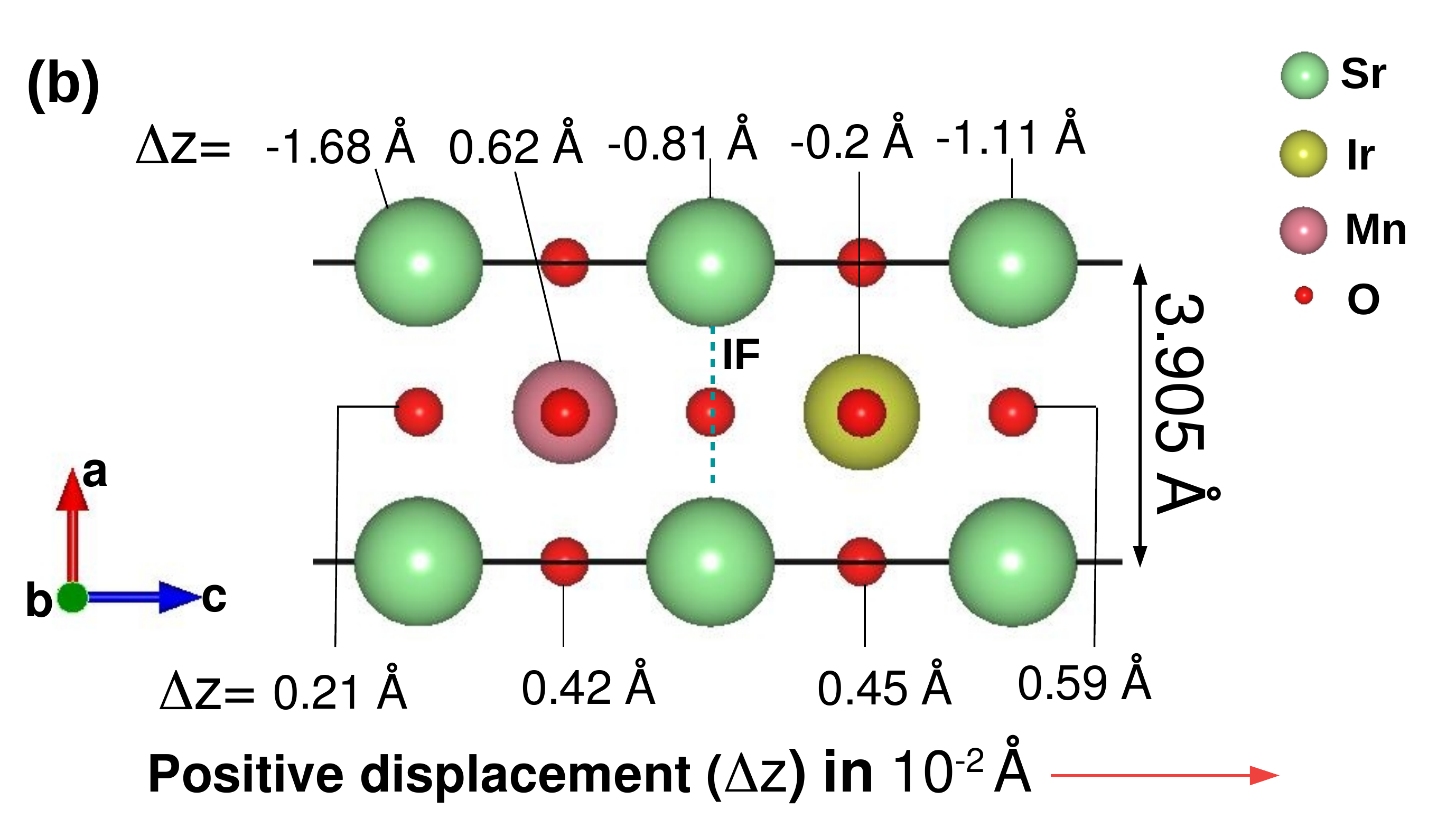}
\caption{Optimized structure for the SIO/SMO slab computed using VASP. Fig. (a) shows the displacements (in \AA) of the atoms in the slab compared to the ideal interface structure. Further atomic displacements (in $10^{-2}$ \AA) in presence of electric field ({\bf E} = 0.3 V/\AA) are shown in (b).} 
\label{fig1}
\end{figure}

The structural optimization of the slab is important to take into account the electrostatic screening effects. In presence of the electric field {\bf E} = 0.3 V/\AA, we can estimate the total ionic dipole moment $\mathcal{D} = \sum_i Z_i \Delta z_i$. Using the atomic displacements $\Delta z_i$ shown in Fig. \ref{fig1} (b) and the formal ionic charges $Z_i$ for simiplicity, this gives a value $\mathcal{D} \sim - 0.12$ e \AA, the sign of which is opposite to the direction of the applied electric field leading to the screening effect. This in turn also affect the AHC.     

In order to show the effect of the electrostatic screening on the AHC, we have computed the AHC for the ideal slab and the relaxed slab  in presence of an electric field {\bf E} (= 0.05 V/\AA).  The calculated AHC for the ideal slab is as large as $\sim$ 78 $\Omega ^{-1}$ cm$^{-1}$, while for the relaxed structure the value is $\sim$ 34 $\Omega ^{-1}$ cm$^{-1}$ indicating the electrostatic screening effects are present in the system.\\

 \section{Model Hamiltonian}
 
This section describes the detail of model Hamiltonian employed in the paper to calculate the Berry curvature and the AHC. 
In the present work, we have studied the interface between two perovskite structures, where the TM elements are arranged on a square lattice. Hence, we have considered a tight-binding model for the $d$ orbitals on a square lattice in presence of an electric field, 
\begin{equation}\label{ham1}
 {\cal H} ={\cal H}_{kin} + {\cal H}_{ex} + {\cal H}_{SOC} + {\cal H}_{E}.
\end{equation}

The terms in the model given in Eq. \ref{ham1} represent the tight-binding Hamiltonian (TBH), exchange splitting to take into account the broken time reversal (TR) symmetry, the atomic SOC, and the external electric field respectively. The spin quantization axis is taken along the ${\hat z}$-direction.  
In the following, we will describe each of the Hamiltonian separately. 

We have used the TBH on a square lattice for the TM-$d$ orbitals ($m$) on site $i$ with the 
field operators $c_{im\sigma}$ and $c^\dagger_{im\sigma}$. In this TBH, hopping upto second nearest neighbor (NN) are considered. The Hamiltonian is written in the Bloch function basis
\begin{equation}\label{Bloch}
 c^\dagger_{\vec{k}m\sigma} = \frac{1}{\sqrt{N}} \sum_{i} e^{i \vec{k}\cdot (\vec{R}_i+\vec{\tau}_m)} c^\dagger_{im\sigma},
\end{equation}
where $\vec{k}$ is the Bloch momentum in the 2D interface BZ. 
The TBH with the order of the basis set: $z^2 \uparrow$, $z^2 \downarrow$, $x^2-y^2 \uparrow$, $x^2-y^2\downarrow$, $xy \uparrow$, $xy \downarrow$, $xz \uparrow$, $xz \downarrow$, $yz \uparrow$ and $yz \downarrow$ is given by,
 
 \begin{eqnarray}\label{H_{kin}}
{H}_{kin} ({\bf k}) = 
\left[ 
{\begin{array}{*{20}c}
   E_1({\bf k})  & \   E_3({\bf k})     & \  E_8({\bf k})   & \ 0            & \ 0\\
   E_3({\bf k})  & \   E_2({\bf k})     & \  0              & \ 0            & \ 0\\
   E_8({\bf k})  & \   0                & \ E_4({\bf k})    & \ 0            & \ 0\\      
   0             & \   0                & \ 0               & \ E_5({\bf k}) & \ E_7({\bf k})\\
   0             & \   0                & \ 0               & \ E_7({\bf k}) & \ E_6({\bf k}) \\
 \end{array} }  \right]
 \otimes 
 \left[
 {\begin{array}{*{20}c}
   1 & \ 0 \\
   0 & \ 1 \\
  \end{array}}\right],
 \end{eqnarray}
 
 where 
 \begin{eqnarray}
  E_1({\bf k}) &=& 2\tilde{t}_1 (\cos k_x + \cos k_y) + 4\tilde{t}_7 \cos k_x\cos k_y + \Delta \nonumber \\
  E_2({\bf k}) &=& 2\tilde{t}_2 (\cos k_x + \cos k_y) + 4\tilde{t}_8 \cos k_x\cos k_y + \Delta \nonumber \\
  E_3({\bf k}) &=& 2\tilde{t}_3 (\cos k_x - \cos k_y) \nonumber \\
  E_4({\bf k}) &=& 2\tilde{t}_4 (\cos k_x + \cos k_y) + 4\tilde{t}_9 \cos k_x\cos k_y \nonumber \\
  E_5({\bf k}) &=& 2\tilde{t}_4 \cos k_x  + 4\tilde{t}_{10} \cos k_x\cos k_y \nonumber \\
  E_6({\bf k}) &=& 2\tilde{t}_4 \cos k_y  + 4\tilde{t}_{10} \cos k_x\cos k_y \nonumber \\
  E_7({\bf k}) &=& 4 \tilde{t}_5 \sin k_x \sin k_y \nonumber \\
  E_8({\bf k}) &=& 4 \tilde{t}_6 \sin k_x \sin k_y . 
 \end{eqnarray}

The parameters of the model ($\tilde{t}_i$, $i = 1, 10$) are obtained from Harrison's table \cite{Harrison}. All the calculations of the paper are performed using $V_\sigma= -0.2 $ eV for the 1NN and  -0.1 eV for the 2NN, $V_\sigma/V_\pi =-1.85$  and $\Delta_{cf} = 3$ eV, where $\Delta_{cf}$ represents the $t_{2g}$-$e_g$ splitting of the $d$-orbitals. The inter-orbital hopping parameters $\tilde{t}_5$ and $\tilde{t}_6$ play the key role in presence of inversion symmetry. 
In absence of these inter-orbital hopping parameters $\tilde{t}_5$ and $\tilde{t}_6$, the Berry curvature vanishes. 

The exchange splitting term in Eq. \ref{ham1}, ${\cal H}_{ex} = - J_{ex} \sum_{i\mu} \sum_{\sigma,\sigma^\prime} c_{i\mu \sigma}^\dagger \sigma_{\sigma\sigma^\prime}^z c_{i\mu\sigma^\prime} $ splits the spin-up and down states. DFT calculations for SIO/SMO interface show that the spins are preferred to align along the $\hat z$ direction in agreement with the experimental results. In view of this, we have considered the direction of the spins to be perpendicular to the square lattice (along the $\hat z$-direction) and hence spin splitting only along that direction is considered. The broken TR symmetry by this term, ensures a non-zero Berry curvature [$\Omega_n({\bf k}) \neq -\Omega_n({-\bf k})$].  

Now, turning to the third term in Eq. \ref{ham1}, the SOC Hamiltonian $ {\cal H}_{SOC} = \lambda {\bm L} \cdot {\bm S}$ has the following form in the above mentioned basis, 

\begin{eqnarray}\label{H_{SOC}}
{\cal H}_{SOC} = \frac{\lambda}{2}
\left[ 
{\begin{array}{*{20}c}
   0  & \ 0    & \ 0   & \ 0 & \ 0 & \ 0 & \ 0 & \  -\sqrt{3} \ & 0 \ & \sqrt{3}i\\
   0  & \ 0    & \ 0   & \ 0 & \ 0 & \ 0 & \ \sqrt{3} & \ 0 \ & \sqrt{3}i & 0 \\
   0  & \ 0    & \ 0   & \ 0 & \ -2i & \ 0 & \ 0 & \ 1 & \ 0 \ & i \\
   0  & \ 0    & \ 0   & \ 0 & \ 0 & \ 2i & \ -1 & \ 0 & \ i & \ 0 \\
   0  & \ 0    & \ 2i  & \ 0 & \ 0 & \ 0 & \ 0 & \ -i & \ 0 & \ 1 \\
   0  & \ 0    & \ 0   & \ -2i & \ 0 & \ 0 & \ -i  & \ 0 & \ -1 & \ 0 \\
   0  & \ \sqrt{3} \ & 0 \ & -1 & \ 0 & \ i & \ 0 & \ 0 & \ -i & \ 0 \\
   -\sqrt{3} & \ 0 & \ 1 & \ 0 & \ i & \ 0 & \ 0 & \ 0 & \ 0 & \ i \\
    0 \ & -\sqrt{3}i & \ 0 & \ -i & \ 0 & \ -1 & \ i & \ 0    & \ 0   & \ 0 \\
    -\sqrt{3}i & \ 0 & \ -i & \ 0 & 1 & \ 0    & \ 0 &\ -i & \ 0    & \ 0
 \end{array} }  \right].
\end{eqnarray}

The SOC deflects the up and down spin in opposite directions which develops a difference in voltage in a spin-polarized system leading to anomalous contribution to the Hall voltage, known as anomalous Hall effect \cite{KL}. 

Finally, we have the electric field term, the last term in Eq. \ref{ham1}. The presence of an external electric field gives rise to additional inter-orbital hopping due to the broken inversion symmetry along the $z$-direction as shown schematically in the main paper (see Fig. 2). These induced inter-orbital hopping parameters are denoted as: $\alpha = \langle xy | H_E | yz\rangle_{\hat x} = \langle xy | H_E | xz\rangle_{\hat y}$, $\beta = \langle xz | H_E | z^2\rangle_{\hat x} =\langle yz | H_E | z^2\rangle_{\hat y}$,
 $\gamma = \langle xz | H_E | x^2 - y^2\rangle_{\hat x} = \langle x^2 - y^2 | H_E  | yz\rangle_{\hat y} $, where the subscript indicates the direction of the nearest neighbor. These hopping parameters are proportional to the applied electric field and due to symmetry reverse their sign as direction of hopping alters. This leads to a sine factor in the Bloch sum as opposed to the cosine factor which gives $k^2$ band dispersion in the TBH. Such sine factors manifests linear $k$ dependence in the band structure. The Hamiltonian in presence of electric field is therefore \cite{Rashba_Shanavas}, 

\begin{eqnarray}\label{H_{E}}
{H}_{E} ({\bf k}) = 2i
\left[ 
{\begin{array}{*{20}c}
   0  & \   0     & \  0   & \ -\beta \sin k_x           & \ -\beta \sin k_y\\
   0  & \   0     & \  0   & \ -\gamma \sin k_x           & \ \gamma \sin k_y\\
   0  & \   0     & \  0   & \ \alpha \sin k_y           & \ \alpha \sin k_x\\
   \beta \sin k_x & \ \gamma \sin k_x  & \  -\alpha \sin k_y & \ 0  & \   0  \\
   \beta \sin k_y & \ -\gamma \sin k_y  & \  -\alpha \sin k_x & \ 0  & \   0  \\
 \end{array} }  \right]
 \otimes 
 \left[
 {\begin{array}{*{20}c}
   1 & \ 0 \\
   0 & \ 1 \\
  \end{array}}\right],
 \end{eqnarray}
 

\section{The Role of symmetry and Inter-Orbital Hopping in Berry Curvature}
 
 In this section we will discuss the role of symmetry in obtaining a non-zero Berry curvature and also show that in absence of inter-orbital hopping the Berry curvature vanishes. The momentum-space Berry-curvature  $\Omega_n({\vec  k})$ for the $n^{\rm th}$ band is a geometric property of the band-structure that manifests the AHC in the system.
 In presence of TR symmetry, $\Omega_n({\vec  k}) = -\Omega_n({-\vec  k})$ while presence of inversion symmetry implies $\Omega_n({\vec  k}) = \Omega_n({-\vec  k})$. Hence in presence of both TR and inversion symmetry Berry curvature becomes zero \cite{MacDonald}. 
 
 In the present case, the magnetization breaks the TR symmetry leading to non-zero $\Omega_n({\vec  k})$ [$\Omega_n({\bf k}) \neq -\Omega_n({-\bf k})$] even in absence of the electric field. For the magnetization along the $z$-direction, the only non-zero component is $\Omega^z_n({\vec  k})$ which is calculated using the Kubo-formula \cite{Thouless},

 \begin{equation}\label{kubo}
 \Omega^z_n ({\bf k}) = \sum_{n^\prime \neq n} \frac {\langle \psi_{n{\bf k}} | \frac{\partial \cal{H}} {\partial {k_x}} | \psi_{n^\prime{\bf k}} \rangle  
                        \langle \psi_{n^\prime{\bf k}} | \frac{\partial \cal{H}} {\partial {k_y}}| \psi_{n{\bf k}} \rangle - (n \leftrightarrow n^\prime)} {(E_{n^\prime}-E_{n})^2},
\end{equation}
 
The off-diagonal matrix elements of the velocity operator, known as anomalous velocity, contributes to the Berry curvature \cite{Kotani}. This emphasizes the crucial role of inter-orbital hopping parameters in defining the non-zero Berry curvature in the system. Thus, in absence of inter-orbital hopping parameters, the Berry curvature vanishes instantaneously. Indeed, the intrinsic AHE, which is directly connected with the Berry curvature, is an inter-band process \cite{KL}. 

In absence of electric field, the $k$-dependence in $\cal{H}$ occurs through $\mathcal{H}_{kin}$. Hence, for diagonal $\mathcal{H}_{kin}$, the velocity operators $\frac{\partial H} {\partial {k_{\eta}}}, \eta = x,y$ are also diagonal in the basis set $|i \rangle$ i.e., $\langle i | \frac{\partial H} {\partial {k_\eta}} | j \rangle = v_{\eta}^i \delta_{ij}$. We will now show that the numerator of the Kubo formula vanishes in absence of off-diagonal inter-orbital hopping. This is true in general and is independent of any specific lattice. 

Expanding the eigen states of $\mathcal{H}$ in its basis sets $|i \rangle$, $|\psi_{n{\bf k}} \rangle = \sum_{i} a^i_{nk} |i \rangle$, the numerator of the Kubo formula can be written as,
\begin{eqnarray}
 && \sum_{n^\prime \neq n} [\langle \psi_{n{\bf k}} | \frac{\partial H} {\partial {k_x}} | \psi_{n^\prime{\bf k}} \rangle \langle \psi_{n^\prime{\bf k}} | \frac{\partial H} {\partial {k_y}}| \psi_{n{\bf k}} \rangle - n \leftrightarrow n^\prime] \nonumber \\
 &=& \sum_{i j} a^{i*}_{nk} a^{i}_{n^\prime k} a^{j*}_{n^\prime k} a^{j}_{nk} v^i_x v^j_y - \sum_{i j} a^{i*}_{n^\prime k} a^{i}_{nk} a^{j*}_{nk} a^{j}_{n^\prime k} v^i_x v^j_y \nonumber \\
 &=& \sum_{i j} (v^i_x v^j_y-v^j_x v^i_y) a^{i*}_{nk} a^{i}_{n^\prime k} a^{j*}_{n^\prime k} a^{j}_{nk} v^i_x v^j_y \nonumber \\
 &=& 0,
\end{eqnarray}
where we have interchanged the dummy indices $i$ and $j$ in the second summation. Thus, in absence of inter-orbital hopping the Berry curvature becomes zero. Hence, in turn, the AHC which is the BZ sum of the Berry curvature for occupied states also vanishes.

\section{DFT results for Berry curvature near the band-crossing point}
In order to understand the sign change of $\sigma_c$ at {\bf E} = 0.3 V/\AA~as shown in Table-I of the paper, in this section we will compare the total Berry curvature [$\Omega^z ({\bf k})$], the sum of the Berry curvature over all the occupied states near the band crossing point $k_c$, obtained from DFT calculation for  
{\bf E} = 0.3 V/\AA ~and {\bf E} = -0.3 V/\AA~respectively. 
\begin{figure}[h]
\centering
\includegraphics[scale=0.4]{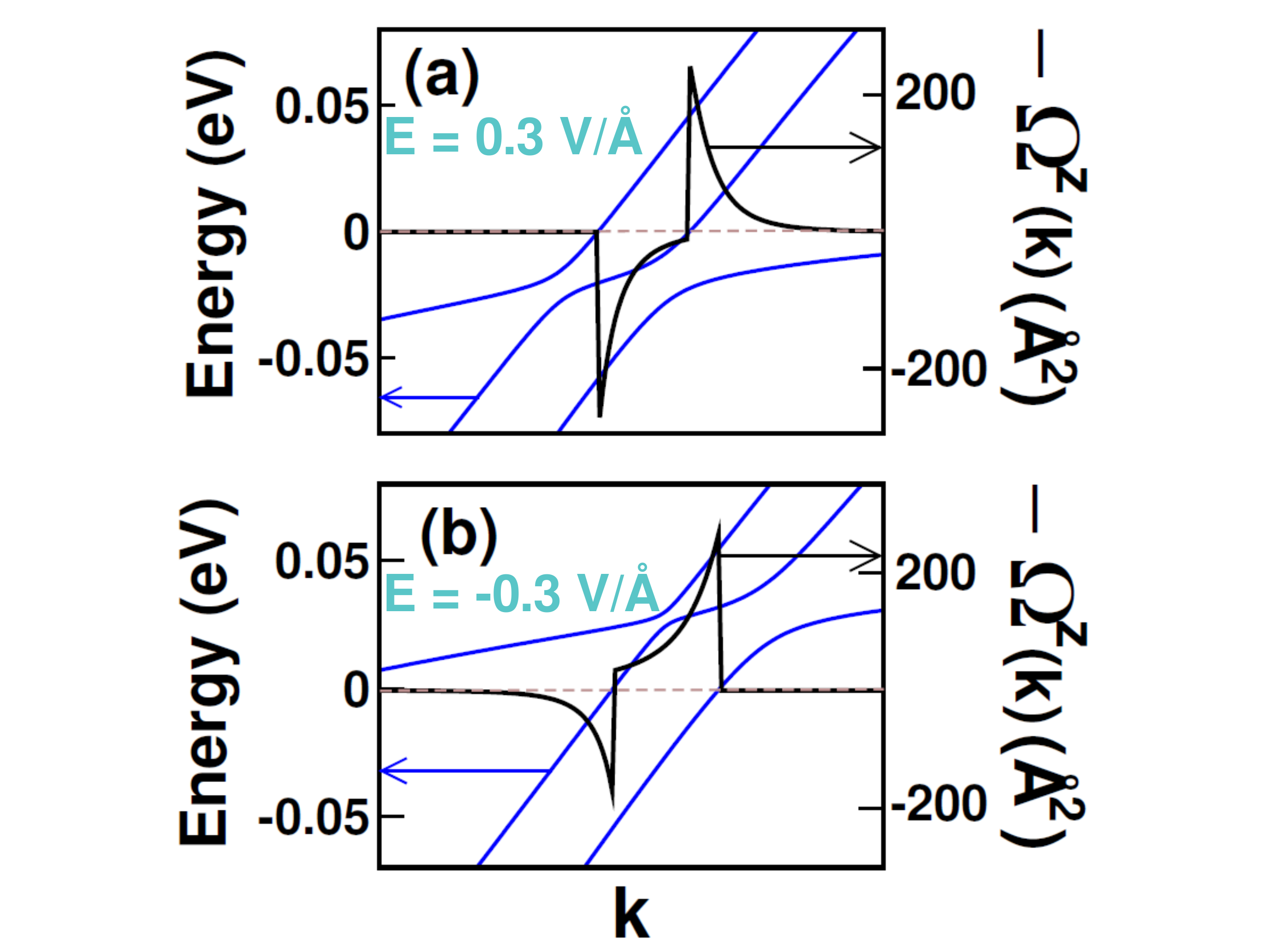}
\caption{The band structure and the total Berry curvature $\Omega^z ({\bf k})$ around the band crossing point for (a) {\bf E} = 0.3 V/\AA ~and (b) {\bf E} = -0.3 V/\AA. It is clear from the figure that bands near the Fermi energy (set at zero, indicated by the dashed line) are modified for {\bf E} = 0.3 V/\AA~ which changes $\Omega^z ({\bf k})$ near the band crossing point.}  
\label{fig3}
\end{figure}

The total Berry curvature $\Omega^z ({\bf k})$ near the band crossing point $k_c$ is shown in Fig. \ref{fig3} (a) and (b) respectively for the applied electric field {\bf E} = 0.3 V/\AA ~and {\bf E} = -0.3 V/\AA. The respective band structures are also shown in the figures. It becomes clear from Fig. \ref{fig3} that the Fermi energy shifts upward for  {\bf E} = 0.3 V/\AA. As a result, both the bands participating in the crossing become occupied partially. As a result of this drastic modification in the band structure near the Fermi level, the total Berry curvature $\Omega^z ({\bf k})$ changes its sign near the crossing point, which is evident from the comparison of Fig. \ref{fig3} (a) and (b) respectively. This explains the sign change of $\sigma_c$ at {\bf E} = 0.3 V/\AA, as listed in Table-I of the paper.

\end{document}